\documentclass[conference]{IEEEtran}
\usepackage{booktabs}
\usepackage{multirow}
\usepackage{graphicx}
\usepackage{amsmath,amssymb}
\usepackage{url}
\usepackage{xcolor}
\usepackage[hidelinks]{hyperref}
\usepackage{adjustbox}
\usepackage{placeins}
\newcommand{\unsafeXVoteMain}{27.8}

\newcommand{\covXVoteMain}{40.6}

\newcommand{\unsafeIndepMain}{16.5}

\newcommand{\unsafePfFiveMain}{1.9}
\newcommand{\unsafeLoPfFiveMain}{1.0}
\newcommand{\unsafeHiPfFiveMain}{2.8}
\newcommand{\covPfFiveMain}{38.2}

\newcommand{\costPfFiveMain}{4.96}

\newcommand{\unsafePfBudgetVoteMain}{12.7}

\newcommand{\covPfBudgetVoteMain}{51.7}

\newcommand{\unsafePfBudgetIndepMain}{5.7}

\newcommand{\faCrossModelSameSourceMain}{62.9}

\newcommand{\faCrossModelIndepSourceMain}{22.9}

\providecommand{\mainActorCalls}{--}

\providecommand{\mainVerifierCallsBase}{--}

\renewcommand{\mainActorCalls}{5,760}

\renewcommand{\mainVerifierCallsBase}{28,362}

\newcommand{\ltUnsafeNoVerifier}{20.8}

\newcommand{\ltUnsafeVoteSame}{9.3}

\newcommand{\ltUnsafeVoteIndep}{5.6}

\newcommand{\ltUnsafeVoteIndepGuardPartialAtomic}{2.8}

\newcommand{\ltSafeVoteIndepGuardPartialAtomic}{41.2}

\newcommand{\ltUnsafePortfolio}{4.2}

\newcommand{\ltSafePortfolio}{68.5}
\newcommand{\ltDeferPortfolio}{16.7}

\newcommand{\ltZeroUpperGuardFullAtomic}{1.4}
\newcommand{\ltSafeGuardFullAtomic}{81.0}

\newcommand{\ltEpisodes}{216}
\newcommand{\ltExecutions}{2016}
\newcommand{\ltPOneDiff}{+0.037}
\newcommand{\ltPOneLo}{+0.009}
\newcommand{\ltPOneHi}{+0.065}
\newcommand{\ltPOneP}{0.2667}
\newcommand{\ltPTwoDiff}{+0.032}
\newcommand{\ltPTwoLo}{+0.014}
\newcommand{\ltPTwoHi}{+0.056}
\newcommand{\ltPTwoP}{0.0960}
\newcommand{\ltPThreeDiff}{+0.014}
\newcommand{\ltPThreeLo}{+0.000}
\newcommand{\ltPThreeHi}{+0.037}
\newcommand{\ltPThreeP}{0.5063}
\newcommand{\ltLatMedVoteSame}{6.09}
\newcommand{\ltLatPnfVoteSame}{7.65}

\newcommand{\ltLatMedNoLoadVoteSame}{2.44}

\newcommand{\ltLatMedVoteIndep}{6.38}
\newcommand{\ltLatPnfVoteIndep}{7.97}

\newcommand{\ltLatMedNoLoadVoteIndep}{2.43}

\newcommand{\ltLatMedPortfolio}{3.44}
\newcommand{\ltLatPnfPortfolio}{4.30}

\newcommand{\ltLatMedNoLoadPortfolio}{1.34}

\newcommand{\bench}{\textsc{VP-Control}}

\newcommand{\Aprime}{A'}

\title{Engineering Reliable Commit Gates for Agentic AI:\\
Cost-Aware Verification Portfolios under Common-Mode Data Failures}

\author{\IEEEauthorblockN{Zihao Zheng\IEEEauthorrefmark{1}, Baichuan Li\IEEEauthorrefmark{2},
Junyi Yao\IEEEauthorrefmark{1}, Jiayu Long\IEEEauthorrefmark{1}}
\IEEEauthorblockA{\IEEEauthorrefmark{1}Washington University in St. Louis, St. Louis, MO, USA}
\IEEEauthorblockA{\IEEEauthorrefmark{2}Southern Methodist University, Dallas, TX, USA}}

\begin{document}
\maketitle

\begin{abstract}
Agentic systems commit state-changing actions, but additional verifiers can
inherit the same upstream fault. We present \bench{}, a runtime-assurance
design and deterministic benchmark for cost-aware commit gates. Its 48 task
templates yield 2,880 scenarios across six fault regimes. A fixed-call
$2{\times}2$ experiment separates verifier-model diversity from evidence-source
diversity. On frozen proposals from two local actor families, a cross-model
vote over shared evidence approves \faCrossModelSameSourceMain\% of unsafe
proposals, versus \faCrossModelIndepSourceMain\% with an independent source.
The source effect is 40.9 percentage points, compared with 11.3 for model
diversity. A portfolio controller selects verification plans using only
deployment-observable metadata. Approximate cluster-adjusted calibration at a
nominal 5\% per-task target yields \unsafePfFiveMain\% unsafe execution and
\covPfFiveMain\% automated safe coverage on the locked test. Matched-budget
portfolios also improve on fixed verification policies. Transfer remains
conditional: unseen fault families yield 16--26\% risk, and a FinQA check fails
to reproduce the source effect with the tested small verifiers. A
preregistered live HTTP/SQLite study tests concurrent writes and lost
responses. After-check races defeat verifier-only gates; transactional partial
guards prevent only covered failures, while a full atomic guard records no
unsafe effects across 216 episodes. Idempotent request identifiers prevent
duplicate effects after lost responses. The results motivate explicit evidence
lineage, cost-aware selection, and commit-time enforcement, while exposing the
limits of approximate calibration and local-tool generalization.

\end{abstract}

\begin{IEEEkeywords}
agentic AI, software engineering, runtime verification, fault injection,
observability, selective automation, risk control
\end{IEEEkeywords}

\section{Introduction}
\label{sec:introduction}

Tool-using language-model agents can restart jobs, modify access control,
promote snapshots, and delete artifacts.  These operations are not merely
answers; they are externally visible software effects.  A production runtime
therefore needs a commit gate between an agent proposal and the target system.
The gate may re-check the proposal, ask another model, read another data source,
run a transactional precondition, or defer to a person.  Each mechanism has a
cost and a distinct failure surface.

The engineering problem is often simplified to ``add another verifier.''  That
rule silently assumes independent evidence.  In deployed data pipelines,
however, two interfaces may be backed by the same replica, cache, extraction
job, or memory record.  A stale upstream can then make different verifier
models agree on the same wrong state.  More votes do not repair a common-mode
data failure. A second boundary arises when correct evidence becomes obsolete
between a check and the write: even an independent read cannot enforce a
precondition that changes before execution. The design question is therefore
which evidence to buy, which clauses to enforce atomically, and when to defer.

Recent work examines multi-agent framework maintenance \cite{liu2025large},
architectural accountability \cite{becattini2025accountability}, and
planner--executor--judge workflows \cite{kaplunovich2025plan}. We study the
commit boundary through an observable gate interface, controlled fault
injection, and comparisons on identical proposals.

We ask four research questions.

\textbf{RQ1: Evidence diversity.}  How much does a second verifier help when it
changes the model, the evidence source, or both, under shared-upstream faults?

\textbf{RQ2: Runtime control.}  Can a controller choose among heterogeneous
verification mechanisms to reduce unsafe commits at a fixed budget, or meet a
risk target without unnecessary deferral?

\textbf{RQ3: Boundaries.}  Which shifts and observability failures break the
result, and does returning better evidence to the actor replace enforcement?

\textbf{RQ4: Live integration.}  When the gate acts on a real tool with
concurrent writes and lost responses, which of these properties survive, and
what do atomic guards and idempotent retries add?

The contribution is an empirical account of these choices at an agent's
commit boundary, with three concrete results. First, a fixed-call factorial
benchmark isolates evidence lineage from model identity: in the controlled
task population, the source effect is substantially larger than the model
effect. Second, an interpretable portfolio uses observable metadata to improve
the risk--cost trade-off over fixed checks on frozen proposals, while exposing
calibration and transfer failures. Third, a live HTTP/SQLite study establishes
the boundary of that benefit: under a forced after-check race, pre-action
evidence is insufficient; a partial atomic guard protects only its declared
clauses, and a full atomic guard is preferable when the entire predicate is
expressible. The benchmark, observable policy interface, and transaction
traces make these comparisons reproducible.

Evidence dependence, atomic checks, and idempotent retries are established
principles. Our contribution lies in their joint evaluation as alternatives
and complements for agent commit gates, including where verification
portfolios cease to help. The main experiment isolates one proposed
state-changing action; the replan study tests one feedback step and the live
arm one action against a real tool, leaving long-horizon coordination outside
the evaluated claim.

\section{Related Work}
\label{sec:related}

\textbf{Verifier dependence.}  Multi-agent verification scales test-time
compute by aggregating verifier outputs \cite{lifshitz2025multi}, and general
LLM-as-a-verifier frameworks broaden the tasks to which this pattern applies
\cite{kwok2026llm}.  Correlated-verifier analyses show that shared errors limit
the reliability gained from a cascade \cite{han2026partially}.  Bara proves the
stronger evidential point that a report-only aggregator cannot distinguish
independent corroboration from duplicated reports descended from one evidence
root \cite{bara2026epistemic}.  We instantiate that distinction at an agent's
commit boundary and vary model and source independently.

\textbf{Source-aware and selective verification.}  ProvenanceGuard routes
claims to source-specific evidence and detects cross-source conflation
\cite{alvarez2026provenanceguard}.  Selective verification can invoke a stronger
model only in an uncertainty region \cite{xue2025verify}; cost-aware retrieval
similarly assigns prices to evidence tiers \cite{wu2026when}.  Sherlock places
verifiers at error-prone workflow nodes under a cost objective
\cite{ro2025sherlock}.  These approaches motivate selection, but they do not
jointly price same-lineage re-checks, independent-source acquisition,
transactional guards, and human deferral under a shared source failure.

\textbf{Risk-controlled action gates.}  Learn-Then-Test (LTT) calibrates a
family of predictive procedures against a risk requirement
\cite{angelopoulos2021learn}; conformal risk control generalizes this view
\cite{angelopoulos2022conformal}.  CORA applies risk-controlled intervention to
mobile GUI agents \cite{feng2026cora}.  Closest to our runtime setting,
SafeCommit chooses probes before a memory-grounded agent commits and otherwise
falls back \cite{akewar2026safecommit}.  Our distinction is architectural:
plans range across mechanisms with different evidence lineages, and the
benchmark explicitly creates source-level common modes.

\textbf{Tool reliability and evaluation.}  ToolGate enforces contracts around
tool execution \cite{liu2026toolgate}; benchmarks study unreliable tool
environments and anomaly recovery \cite{tian2026beyond,zhu2026when}.  Static
replay can be misleading when replacing an upstream component changes the
trajectory \cite{gonuguntla2026replay}.  Our replay is limited to a gate after a
frozen proposal; a separate replan experiment measures the case in which new
evidence changes the actor's next step.

\section{Commit-Gate Design}
\label{sec:design}

\subsection{System model}

A task template $\tau$ defines an operation, a precondition predicate
$\Phi_\tau$, and an effect $E_\tau$.  The actor reads state $s_0$ and proposes
an action $a$ or abstains.  Immediately before execution, the authoritative
state is $s^*$.  A commit is unsafe when $\Phi_\tau(s^*)$ is false or when the
effect already holds for a non-idempotent operation.  The gate returns
\texttt{commit}, \texttt{block}, or \texttt{defer}; only \texttt{commit} invokes
the target system.

The runtime exposes three evidence views.  $A$ is the actor's original read.
$\Aprime$ is a second interface backed by the same upstream lineage as $A$.
$B$ is an independently replicated snapshot with its own lag process.  A
verifier model reads exactly one view and returns approve/reject plus a
confidence.  An exact guard checks the subset of $\Phi_\tau$ expressible by the
target system at commit time.  Deferral models a human decision.  Each view
and mechanism has a unit cost.

\textbf{Illustrative data-pipeline case.}  The
\texttt{apply\_schema\_migration} template requires safe compatibility and
approval before moving the schema from version 4 to 5; its guard covers
compatibility and duplicate execution, not approval.  If approval is revoked
while $A$ and $\Aprime$ retain the old value, two models can approve from the
same stale evidence and a guard pass leaves approval unchecked; only a fresh
$B$ exposes the revocation.  Evidence provenance and guard expressiveness must
therefore be part of the interface rather than inferred from the number of
checks.

A \texttt{ControllerView} interface separates deployment observations from
scoring state. It exposes the proposal, source identifiers and lineage, ages,
noisy health flags, schema completeness, guard availability, blast radius,
reversibility, last tool status, and purchased verdicts. It excludes $s^*$,
the injected fault regime, true source staleness, and the unsafe label.
Static scans check policy code for hidden fields; dynamic audits reconstruct
decisions from the serialized observable view.

\subsection{Verification plans}

We enumerate 37 small decision trees over five mechanism families: same-model
and cross-model re-checks on $\Aprime$, reads of $B$, a transactional guard,
and deferral.  The library includes single checks, two-call unanimity votes,
three same-source resamples, cheap-first escalation, and guard-then-source
plans.  Leaves commit, block, or defer.  A repeated mechanism is charged once,
and fixed baselines are members of the same library.

Observable contexts are buckets over the age of $A$, last-tool-status family,
health of $\Aprime$ and $B$, and guard availability.  They form a six-level
back-off hierarchy.  A level is used only with at least 24 training rows from
three templates; otherwise the controller backs off to a coarser context.
This keeps the learned object an auditable lookup table rather than an opaque
policy model.

For context $x$ and risk price $\nu=1/\lambda$, training selects
\begin{equation}
 p_x(\lambda)=\arg\min_{p\in\mathcal P}
 \bigl[c(p)-w\,q(p)+\nu\,\widehat r(p)\bigr],
 \label{eq:objective}
\end{equation}
where $c$ is total cost, $q$ is automated safe coverage,
$\widehat r=(k+1/2)/(n+1)$ is a smoothed unsafe-commit rate, and $w=4$.
Varying $\lambda$ yields a conservative-to-permissive family of policies.

\subsection{Calibration and budget variants}

On held-out calibration templates, policies are tested in increasing
$\lambda$ order.  The test statistic is a Wilson upper bound on unsafe-commit
risk with effective sample size $n/\mathrm{deff}$; the Kish design effect
$\mathrm{deff}=1+(\bar m-1)\rho$ uses an ANOVA estimate of within-template
correlation.  The selector follows the fixed-sequence organization of LTT:
it stops at the first failing policy and deploys the last passing one
\cite{angelopoulos2021learn}.  If no policy passes, the controller defers
everywhere.  The implementation uses $z=1.96$ and estimates correlation from
the calibration data.  This design-effect-adjusted Wilson bound is an
approximation; we do not establish super-uniform tests for clustered outcomes.
Consequently, the nominal targets are empirical calibration settings, not
finite-sample, distribution-free guarantees.  Exchangeable deployment
templates are necessary for interpreting transfer, but do not by themselves
make this approximate bound exact.

For matched-cost comparisons, each context instead minimizes
$\widehat r-0.04q+\mu c$; bisection chooses $\mu$ so mean training cost does not
exceed the baseline budget.  These variants answer a different question from
the calibrated risk target and carry no risk guarantee.

\section{Implementation and Traceability}
\label{sec:implementation}

The reference implementation separates generation, decision logic, execution,
and analysis so the experimental boundary is visible in code.  Task templates
declare predicates, effects, guard clauses, blast radius, and reversibility.
The world generator materializes the authoritative and rendered states.  A
\texttt{ControllerView} constructor removes hidden fields before policy code is
called.  The controller maps this view and a proposal to a named plan; a plan
executor then resolves verifier roles, executes the decision tree, memoizes
duplicate mechanism calls, charges their costs, and invokes the simulator only
on a commit leaf.

Every decision serializes the selected plan, ordered mechanism steps, concrete
model/source/seed binding, verifier outcome and confidence, verification and
escalation cost, final disposition, and scored outcome.  These records provide
an audit trail from a table entry back to the proposal and raw verdict that
produced it.  Actor and verifier runners are append-only and resumable.  Cache
keys include the model, scenario, source, normalized proposal, seed, and prompt
variant, preventing a verdict for one evidence view or prompt from being reused
as another.

Design choices were separated from final evaluation by stage: Stage~0 used
rule actors/verifiers for feasibility and power, where one parameter retuning
and the move to a global Lagrangian occurred and are logged as preregistration
deviations; Stage~1 used training templates to repair the verifier prompt's
version-equality heuristic; Stage~2 froze proposals, verdicts, and the locked
test; later stages replayed frozen records or created explicitly labeled
stress, external, and live datasets without revising the main policy after
seeing test outcomes.

Two audits check the end-to-end trace.  The static leakage audit searches every
policy-facing function for forbidden hidden fields.  The dynamic audit rebuilds
the observable view from serialized records and requires the same plan choice.
A separate recomputation audit sampled 900 plan entries and recomputed all 33
reported strategy rates from replay rows; it found zero mismatches.  Paper tables
and numeric macros are generated from analysis JSON rather than manually copied
into prose.  These checks do not establish external validity, but they make
policy comparison and reported numbers mechanically inspectable.

\begin{table*}[tp]
\centering\small
\caption{Engineering requirements, evaluated mechanisms, and remaining limits.}
\label{tab:requirements}
\setlength{\tabcolsep}{4pt}
\begin{tabular}{p{0.19\textwidth}p{0.30\textwidth}p{0.43\textwidth}}
\toprule
Requirement & Mechanism and evidence & Limit on the supported claim \\
\midrule
Separate failure domains & Explicit source lineage; fixed-call $2{\times}2$ comparison (RQ1) & Lineage is supplied; independent sources can still lag or share an unmodeled fault \\
Balance safety and availability & Contextual plans; matched-budget and nominal-target evaluation (RQ2) & Unit costs are modeled; calibration bound is approximate \\
Make decisions inspectable & Observable-view boundary, serialized steps, replay recomputation, transaction log with linearization-point pre-state & Audits check consistency and one concurrent writer, not malicious writers \\
Enforce action disposition & Plan leaves gate simulated and live execution; atomic guard inside the write transaction; idempotent retries; bounded replan (RQ3, RQ4) & Local loopback service and SQLite only; one writer process; no long-horizon rollout, distributed failure, or adversarial writer \\
Detect limits of transfer & Domain/fault holdouts and external FinQA check (RQ3) & No automatic deployment drift detector is implemented \\
\bottomrule
\end{tabular}
\end{table*}

\section{Evaluation Method}
\label{sec:evaluation}

\subsection{Controlled benchmark}

\bench{} contains 48 semantic templates, twelve each for schedulers,
pipelines, storage systems, and identity/access management.  Templates specify
an operation, named preconditions, effect, blast radius, reversibility, and the
clauses available to a transactional guard.  Thirty-nine templates support a
guard; policy and integrity clauses are intentionally outside its scope.

Ten hash-seeded worlds per template are instantiated in six regimes:
\emph{clean}; \emph{a\_stale}, in which only $A$ is stale;
\emph{common\_stale}, in which $A$ and $\Aprime$ share a stale upstream;
\emph{partial\_update}; \emph{ambiguous\_commit}; and
\emph{all\_degraded}, in which every source is stale and the guard is usually
unavailable.  Non-clean regimes contain a hazardous latent change with
probability 0.5 and a benign change otherwise, so regime does not reveal the
label.  $B$ independently lags with probability 0.10; source health and schema
signals are noisy.  The resulting $48\times10\times6=2{,}880$ scenarios are
split by template into 24/8/16 train/calibration/test groups, stratified by
domain and ordered by a salted hash.

The $2{\times}2$ experiment begins with $V_1$ on $\Aprime$ and adds one of:
a new seed of $V_1$ on $\Aprime$, $V_2$ on $\Aprime$, $V_1$ on $B$, or $V_2$
on $B$.  Thus every cell uses two calls and separates same/cross model from
same/independent source.  We report false-approve rate on unsafe proposals,
error correlation, and the ratio of observed joint errors to the product of
marginal errors.

\subsection{Models, locking, and counterfactual replay}

\textbf{Metric denominators.}  For $N$ evaluated proposal records with
$U_i$ an unsafe execution, $S_i$ a safely satisfied goal, and $D_i$ deferral,
policy risk is $\sum_i U_i/N$ and safe automated coverage is
$\sum_i S_i(1-D_i)/N$; coverage can include a blocked duplicate when the goal
already holds.  Verifier false-approve rate instead conditions on proposals
that would be unsafe if executed.  A 5\% target therefore concerns unsafe
executions per task, not per executed action, which is why coverage and cost
are reported alongside it.  Cost is a modeled unit charge, not latency or
tokens.

Actors are Qwen3.5-9B and Gemma4-8B; verifiers are Qwen3-4B ($V_1$) and
Phi-4-mini ($V_2$).  All are quantized public Ollama checkpoints run locally,
one resident model at a time.  Actors produce a constrained action schema;
verifiers receive the task contract, proposal, and exactly one rendered source
and produce a constrained verdict schema.  A pilot on training templates
revealed that both initial verifier prompts treated any version change as a
failure.  Before the main run, the prompt was changed once to state that a
version difference alone is not grounds for rejection; the change and both
pilot outputs are retained.  Fault rates, costs, plans, splits, and statistical
procedures were then locked.

\begin{table}[t]
\centering\small
\caption{Frozen public checkpoints.  Digest prefixes identify the model blobs.}
\label{tab:models}
\adjustbox{max width=\linewidth}{\begin{tabular}{llllll}
\toprule
Model & Role & Params & Quant. & Registry digest & Ctx \\
\midrule
Qwen3.5-9B & actor & 9.7B & Q4\_K\_M & \texttt{6488c96fa5fa} & 262144 \\
Gemma4-8B & actor & 8.0B & Q4\_K\_M & \texttt{c6eb396dbd59} & 131072 \\
Qwen3-4B & verifier $V_1$ & 4.0B & Q4\_K\_M & \texttt{359d7dd4bcda} & 262144 \\
Phi-4-mini & verifier $V_2$ & 3.8B & Q4\_K\_M & \texttt{78fad5d182a7} & 131072 \\
Llama-3.2-3B & weak verifier $V_w$ (stress) & 3.2B & Q4\_K\_M & \texttt{a80c4f17acd5} & 131072 \\
\bottomrule
\end{tabular}
}
\end{table}

The main run freezes \mainActorCalls{} proposals and caches
\mainVerifierCallsBase{} verifier calls.  All outputs are valid JSON and no
main-run actor abstains.  Every controller is replayed over these identical
proposals and cached verdicts.  A second arm uses rule-based actors and
verifiers whose idiosyncratic errors are independent across models; it tests
whether the result follows from the evidence structure rather than a specific
LLM family.

The template is the unit of analysis.  Confidence intervals use 2,000 cluster
bootstrap samples.  Paired differences use 5,000 template-cluster sign-flip
permutations.  Three primary comparisons were fixed in advance and corrected
with Holm's method; other analyses are exploratory.  Risk targets are 1\%,
2\%, and 5\%.

\subsection{Boundary checks}

We refit while holding out each domain or fault family, shift verifier
prompts, replace one verifier with a weaker model, swap actor/verifier roles,
vary seeds and unit costs, and allow one bounded actor replan after receiving
$\Aprime$ or $B$.  A 240-question FinQA subset \cite{chen2021finqa} yields 960
document-grounded scenarios with executable gold programs and controlled
table corruption, used only as an external check.

\subsection{Live tool-integration arm}
\label{sec:live-method}

The simulator evaluates a fixed commit state.  To test the integration
requirements of Section~\ref{sec:implications} on real software, we add a
preregistered local arm in which the gate acts on a live tool: an HTTP service
bound to the loopback interface over a per-episode SQLite database
(rollback journal, synchronous writes, bounded lock wait), a separate writer
process, and the same local checkpoints.  Twelve business contracts cover three
operation families---schema migration (\texttt{ALTER TABLE} plus a migration
ledger), snapshot promotion (pointer switch plus audit row), and failed-job
restart (attempt row plus status, with a unique index on active attempts)---in
four variants that change the declared predicate or the guard's coverage.  The
service exposes views $A$ and $\Aprime$ from one shared cache file, a view $B$
that reads the database through a fresh connection, a non-atomic
\texttt{preflight}, and a \texttt{commit} whose guard re-reads its covered
clauses inside the same \texttt{BEGIN IMMEDIATE} transaction that applies the
effect.  A partial guard covers mechanical state and duplicate effects but not
approval; a full guard covers the whole predicate.  Duplicate columns and
duplicate active attempts are refused by the database itself and are recorded
as tool errors, not converted into successes.

Six conditions are injected per contract and seed with hash-assigned
hazard/benign branches: C0 clean; C1 benign change; C2 common stale cache
(approval revoked, validation broken, effect already applied, or an extra
precondition broken after the cache was built, while $B$ sees the change); C3
after-check race (the writer process commits between the gate's last check and
the action, synchronised by a barrier); C4 ambiguous response (the service
commits, then closes the connection without answering; the client retries with
the same request identifier); and C5 degraded evidence (stale cache and $B$
times out).  Cache age is the real number of committed transactions since the
snapshot.  Eight policies run from an identical database copy per episode: plain
commit; cross-model votes on the shared cache ($\Aprime{+}\Aprime$) or with an
independent read ($\Aprime{+}B$); partial guard as non-atomic preflight and as
atomic check; the independent vote followed by the atomic partial guard; the
frozen 5\% portfolio table from Section~\ref{sec:results}, evaluated zero-shot
through a fixed metadata mapping with the atomic partial guard as its guard
adapter; and the full atomic guard.  Unsafe execution is scored from the
predicate at the transaction's linearization point, which the plain commit path
records but never acts on.  Three comparisons were fixed before the run and
Holm-corrected: L1 shared-cache vote versus independent-read vote, L2 preflight
versus atomic partial guard, and L3 portfolio versus independent vote with the
same atomic guard.  Layer~A reuses verifier answers cached by evidence content
so that policies are compared on identical verdicts while every database effect
is executed; layer~B repeats a 72-episode subset with every model call live and
one resident model at a time, reporting post-proposal gate latency with and
without model load time.

\section{Results}
\label{sec:results}

\subsection{RQ1: Source diversity has the larger effect in the benchmark}

Table~\ref{tab:2x2} reports the two-call unanimity vote on unsafe locked-test
proposals.  Pooled over fault regimes, a second seed of the same model on the
same source approves 74.2\% of unsafe proposals.  Changing only the verifier
family reduces this to \faCrossModelSameSourceMain\%.  Holding the model fixed
and changing to an independent source reduces the rate to 33.3\%; changing
both reaches \faCrossModelIndepSourceMain\%.

\begin{table}[t]
\centering\small
\caption{Single verifier calls on executable locked-test proposals (\%).}
\label{tab:verifiers}
\adjustbox{max width=\linewidth}{\begin{tabular}{lrrrrr}
\toprule
Verifier call & Accuracy & False approve & False reject & AUROC & ECE \\
\midrule
$V_1$ on $A'$ & 60.1 & 75.0 & 12.1 & 0.555 & 0.373 \\
$V_1$ on $A'$ (seed 2) & 59.6 & 75.4 & 12.6 & 0.588 & 0.367 \\
$V_2$ on $A'$ & 54.4 & 63.6 & 31.4 & 0.527 & 0.407 \\
$V_1$ on $B$ & 68.7 & 37.3 & 26.5 & 0.666 & 0.278 \\
$V_2$ on $B$ & 53.7 & 25.4 & 62.8 & 0.542 & 0.309 \\
\bottomrule
\end{tabular}
}
\end{table}

\begin{table}[t]
\centering\small
\caption{False-approve rate (\%) of two-call unanimity votes on unsafe
locked-test proposals.  Columns are same/cross model by same/independent
source.}
\label{tab:2x2}
\adjustbox{max width=\linewidth}{\begin{tabular}{lrrrrr}
\toprule
Regime & same/same & cross/same & same/indep & cross/indep & $n$ unsafe \\
\midrule
a stale & 16.3 & 12.2 & 2.9 & 2.9 & 172 \\
all degraded & 87.4 & 75.3 & 80.5 & 41.4 & 174 \\
ambiguous commit & 95.0 & 86.2 & 61.3 & 55.0 & 160 \\
common stale & 91.5 & 81.7 & 9.8 & 7.3 & 164 \\
partial update & 82.7 & 61.5 & 13.4 & 9.5 & 179 \\
\midrule
pooled & 74.2 & 62.9 & 33.3 & 22.9 & 849 \\
\bottomrule
\end{tabular}
}
\end{table}

\begin{figure}[t]
\centering
\includegraphics[width=\linewidth]{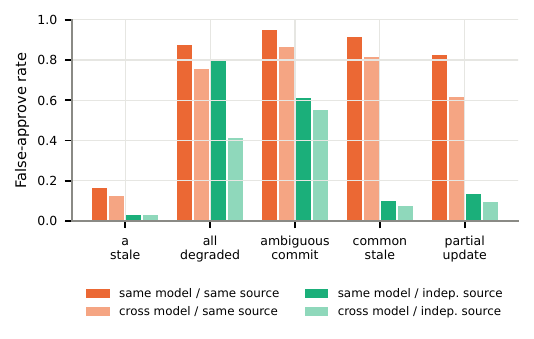}
\caption{The same $2{\times}2$ comparison by fault regime.  Independent
evidence is most valuable when the shared lineage is stale and loses that
advantage when all sources degrade.}
\label{fig:2x2-regime}
\end{figure}

The paired source effect at fixed model is -0.409
[-0.479, -0.331], versus -0.113 [-0.215, -0.038] for the model effect at
fixed source.  The difference-in-differences is 0.008
[-0.064, 0.101] ($p=0.9056$), so we do not detect an interaction.  The regime
rows identify the mechanism: under \emph{common\_stale}, cross-model approval
falls from 81.7\% on the shared source to 7.3\% when one call uses $B$; under
\emph{all\_degraded}, even the cross-model independent-source cell remains at
41.4\%.  An independent read is useful only when it has an independent failure
path.

The rule-based arm reaches the same qualitative result even though model-side
errors are independent by construction: cross-model approval is 65.9\% with a
shared source and 22.8\% with an independent source.  This isolates the main
effect in the evidence plumbing rather than in one pair of LLMs.

\subsection{RQ2: A portfolio improves the risk--cost trade-off}

Table~\ref{tab:strategies} keeps the baselines needed to interpret the
controller.  Always executing is unsafe on 44.2\% of scenarios.  A cross-model
same-source vote lowers this to \unsafeXVoteMain\%, but an independent source
does more (\unsafeIndepMain\%).  Guard-then-$B$ reaches 0.8\% by deferring on
56.2\% of scenarios and costs 8.82 units.

\begin{table*}[tp]
\centering\small
\caption{Locked-test policies (LLM arm).  Unsafe, coverage, and defer
are percentages; total cost uses prices $\Aprime=1$, $B=3$, guard $=2$, and
human deferral $=10$.}
\label{tab:strategies}
\adjustbox{max width=\textwidth}{\begin{tabular}{lrrrrr}
\toprule
Strategy & Unsafe \% [95\% CI] & Coverage \% & Defer \% & Verif. cost & Total cost \\
\midrule
Always execute & 44.2 [40.8, 47.5] & 55.7 & 0.0 & 0.00 & 0.00 \\
Same-source check ($V_1$ on $A'$) & 33.2 [26.9, 38.3] & 50.3 & 0.0 & 1.00 & 1.00 \\
Majority of 3 resamples ($V_1$ on $A'$) & 33.3 [27.0, 38.5] & 50.0 & 0.0 & 3.00 & 3.00 \\
Cross-model vote, same source & 27.8 [20.6, 34.5] & 40.6 & 0.0 & 2.00 & 2.00 \\
Same-model vote, $A'$+$B$ & 14.7 [11.0, 18.5] & 45.3 & 0.0 & 4.00 & 4.00 \\
Cross-model vote, $A'$+$B$ & 10.1 [6.8, 13.8] & 27.5 & 0.0 & 4.00 & 4.00 \\
Independent source ($V_1$ on $B$) & 16.5 [13.0, 20.0] & 45.6 & 0.0 & 3.00 & 3.00 \\
Independent source ($V_2$ on $B$) & 11.2 [7.8, 14.9] & 27.2 & 0.0 & 3.00 & 3.00 \\
Both models on $B$ & 10.2 [6.9, 13.8] & 27.6 & 0.0 & 6.00 & 6.00 \\
Exact guard (defer if unavailable) & 6.7 [4.1, 9.4] & 40.7 & 41.0 & 2.00 & 6.10 \\
Guard, then $V_1$ on $B$ & 0.8 [0.2, 1.4] & 31.4 & 56.2 & 3.20 & 8.82 \\
Always defer & 0.0 [$\le$0.2] & 0.0 & 99.9 & 0.00 & 9.99 \\
TTL gate & 19.6 [16.6, 22.6] & 49.2 & 0.0 & 2.21 & 2.21 \\
Confidence gate & 44.2 [40.8, 47.5] & 55.7 & 0.0 & 0.00 & 0.00 \\
Evidence router baseline & 28.6 [24.4, 32.7] & 52.6 & 0.0 & 1.36 & 1.36 \\
\midrule
Random mechanism (matched cost) & 20.9 [16.1, 25.5] & 41.8 & 0.0 & 2.02 & 2.02 \\
\midrule
Global LTT @1\% & 0.0 [$\le$0.2] & 0.0 & 99.9 & 0.00 & 9.99 \\
Global LTT @2\% & 0.8 [0.2, 1.4] & 31.4 & 56.2 & 3.20 & 8.82 \\
Global LTT @5\% & 0.8 [0.2, 1.4] & 31.4 & 56.2 & 3.20 & 8.82 \\
Confidence-threshold LTT @1\% & 0.0 [$\le$0.2] & 13.4 & 0.0 & 2.00 & 2.00 \\
Confidence-threshold LTT @2\% & 0.0 [$\le$0.2] & 13.4 & 0.0 & 2.00 & 2.00 \\
Confidence-threshold LTT @5\% & 0.0 [$\le$0.2] & 13.4 & 0.0 & 2.00 & 2.00 \\
\midrule
\textbf{Portfolio @1\%} & 0.0 [$\le$0.2] & 0.0 & 99.9 & 0.00 & 9.99 \\
\textbf{Portfolio @2\%} & 0.0 [$\le$0.2] & 0.0 & 99.9 & 0.00 & 9.99 \\
\textbf{Portfolio @5\%} & 1.9 [1.0, 2.8] & 38.2 & 31.4 & 1.83 & 4.96 \\
\midrule
\textbf{Portfolio, budget of cross-model vote} & 12.7 [9.5, 16.4] & 51.7 & 2.1 & 1.78 & 1.99 \\
\textbf{Portfolio, budget of independent source} & 5.7 [3.6, 8.4] & 45.3 & 16.6 & 1.72 & 3.37 \\
\textbf{Portfolio, budget of exact guard} & 2.6 [1.5, 4.0] & 39.6 & 29.9 & 1.82 & 4.82 \\
\textbf{Portfolio, risk of independent source} & 24.7 [19.7, 30.1] & 58.1 & 0.0 & 0.88 & 0.88 \\
\textbf{Portfolio, risk of cross-model vote} & 29.9 [25.4, 34.7] & 59.7 & 0.0 & 0.52 & 0.52 \\
\bottomrule
\end{tabular}
}
\end{table*}

\begin{figure*}[tp]
\centering
\includegraphics[width=0.92\textwidth]{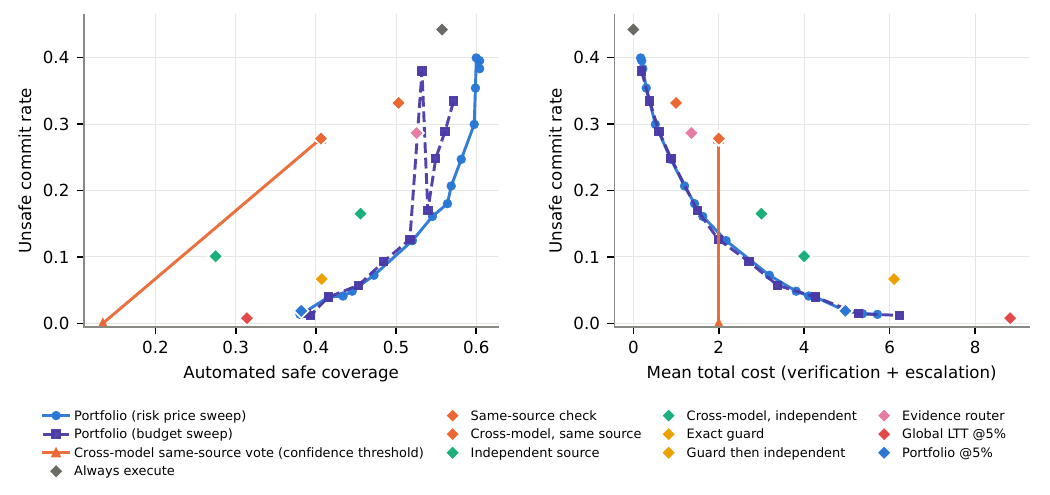}
\caption{Locked-test risk--coverage and risk--cost planes.  Curves sweep the
portfolio families; diamonds are fixed strategies.  The figure reports the
trade-off rather than collapsing risk, availability, and cost into one score.}
\label{fig:frontier}
\end{figure*}

The 5\% calibration selects a portfolio whose locked-test unsafe rate is
\unsafePfFiveMain\% [\unsafeLoPfFiveMain, \unsafeHiPfFiveMain], with
\covPfFiveMain\% safe automated coverage and cost \costPfFiveMain.  Its
calibration result uses an approximate cluster-adjusted bound; the observed
test rate is empirical evidence, not a finite-sample safety certificate.
The 1\% and 2\% targets cannot be resolved with eight calibration templates;
the controller therefore defers essentially everywhere.

At the cross-model vote's budget, the portfolio reduces unsafe commits from
\unsafeXVoteMain\% to \unsafePfBudgetVoteMain\% while increasing coverage from
\covXVoteMain\% to \covPfBudgetVoteMain\%.  At the independent-source budget,
it reduces unsafe commits from \unsafeIndepMain\% to
\unsafePfBudgetIndepMain\% at nearly unchanged coverage.  Table~\ref{tab:primary}
shows the three preregistered paired comparisons; all remain significant after
Holm correction.  The simulated arm agrees: at the vote budget it yields 11.9\%
versus 25.3\% unsafe, and at the independent-source budget 6.0\% versus 11.9\%.

\begin{table}[t]
\centering\small
\caption{Primary paired comparisons.  Negative differences favor the
portfolio or independent source.}
\label{tab:primary}
\adjustbox{max width=\linewidth}{\begin{tabular}{lrr}
\toprule
Preregistered comparison & Difference [95\% CI] & Holm $p$ \\
\midrule
P1: portfolio (vote budget) $-$ vote & -0.152 [-0.209, -0.092] & 0.0006 \\
P2: portfolio (indep. budget) $-$ indep. & -0.108 [-0.140, -0.076] & 0.0006 \\
P3: indep. source $-$ same source & -0.409 [-0.479, -0.331] & 0.0006 \\
\bottomrule
\end{tabular}
}
\end{table}

\subsection{RQ3: Transfer is conditional and enforcement remains necessary}

Domain holdouts are mixed.  The 5\% portfolio remains within target on
scheduler and storage, but cannot be calibrated on IAM and pipeline and defers
throughout those folds.  The budget-matched variant transfers in all four
domains (4.6--7.0\% unsafe versus 14.5--18.5\% for the independent-source
baseline at equal cost).  Fault-family holdouts are harsher: a controller that
has not seen \emph{ambiguous\_commit} or \emph{partial\_update} reaches 26\%
and 16\% realised risk, respectively, at the nominal 5\% target.  These folds
intentionally violate exchangeability and demonstrate a failure to maintain
the nominal target.

\begin{table*}[tp]
\centering\small
\caption{Out-of-distribution refits.  Each entry is unsafe \% / safe automated
coverage \%.}
\label{tab:ood}
\adjustbox{max width=\textwidth}{\begin{tabular}{lrrrrr}
\toprule
Held-out fold & Indep.\ source & Cross-model vote & Global LTT@5\% & Portfolio@5\% & Portfolio, indep.\ budget \\
\midrule
\multicolumn{6}{l}{\emph{unseen domain} (unsafe \% / coverage \%)} \\
\quad iam & 18.5 / 46.3 & 26.2 / 41.6 & 0.8 / 30.0 & 0.0 / 0.0 & 4.6 / 44.0 \\
\quad pipeline & 14.5 / 50.2 & 23.8 / 41.6 & 1.5 / 47.0 & 0.0 / 0.0 & 6.9 / 50.3 \\
\quad scheduler & 17.3 / 52.8 & 29.7 / 49.2 & 0.7 / 43.7 & 1.0 / 43.8 & 6.7 / 53.8 \\
\quad storage & 16.5 / 46.1 & 21.7 / 38.0 & 1.5 / 38.5 & 2.8 / 45.5 & 7.0 / 48.8 \\
\multicolumn{6}{l}{\emph{unseen regime} (unsafe \% / coverage \%)} \\
\quad a stale & 6.9 / 34.4 & 6.6 / 19.1 & 1.2 / 24.4 & 4.1 / 25.0 & 9.7 / 30.0 \\
\quad all degraded & 47.5 / 37.5 & 40.9 / 35.3 & 2.2 / 4.4 & 4.1 / 7.5 & 10.3 / 13.8 \\
\quad ambiguous commit & 31.9 / 55.3 & 43.1 / 44.7 & 0.0 / 58.4 & 26.2 / 55.3 & 27.5 / 55.3 \\
\quad clean & 0.0 / 89.1 & 0.0 / 77.2 & 0.0 / 59.1 & 0.0 / 50.9 & 0.0 / 90.6 \\
\quad common stale & 5.0 / 34.1 & 41.9 / 37.5 & 18.1 / 30.6 & 0.0 / 0.0 & 30.0 / 40.3 \\
\quad partial update & 7.8 / 23.1 & 34.4 / 30.0 & 0.0 / 18.1 & 16.2 / 21.6 & 34.1 / 32.2 \\
\bottomrule
\end{tabular}
}
\end{table*}

Under stress conditions (verifier-changing rows refit, others reuse the
learned table), moderate verifier-context shift leaves the 5\% portfolio at 2.2\% unsafe, and
substituting a weak verifier yields 2.8\% after refitting.  Swapping the small
and large model roles yields 2.1\%.  Varying independent-read and deferral
prices keeps the risk-controlled policy between 1.7\% and 2.6\% while changing
its mechanism mix.  In contrast, pooling additional actor seeds makes the 5\%
candidate family fail calibration, so it defers rather than extrapolating.

Returning evidence to the actor does not replace the gate.  Among 192 initially
unsafe proposals in a pre-fixed 25\% test subset with one bounded replan, a same-lineage view leaves
92.2\% unsafe after one replan; $B$ lowers this to 69.8\%, but the actor abstains
only 30.2\%.  The independent evidence helps, yet most unsafe proposals would
still be recommitted without enforcement, consistent with reports that agents
may fail to act on known risk \cite{tang2025lm}.

The FinQA check is a negative external result.  The actor is already wrong
on 43.3\% of uncorrupted tables; Phi-4-mini rejects every answer, and Qwen3-4B
approves 17\% of unsafe answers from the corrupted shared rendering versus
20\% from the correct independent table, reversing the source effect.  The
zero-shot portfolio reaches 5.1\% unsafe at 9.4\% coverage, mostly through
the executable guard and deferral.  Source independence matters only once a
verifier is competent for the evidence and task, so we make no cross-domain
claim.

\begin{table}[t]
\centering\small
\caption{External FinQA check (\%, except cost).  The portfolio is transferred
without selecting parameters on FinQA.}
\label{tab:external}
\adjustbox{max width=\linewidth}{\begin{tabular}{lrrrr}
\toprule
Policy & Unsafe & Coverage & Defer & Cost \\
\midrule
Always submit & 56.7 & 30.8 & 0.0 & 0.00 \\
Same-source check & 12.8 & 13.9 & 0.0 & 0.88 \\
Independent source & 18.4 & 18.0 & 0.0 & 2.62 \\
Executable guard & 6.1 & 20.9 & 18.9 & 3.64 \\
Portfolio @5\% (zero-shot) & 5.1 & 9.4 & 20.7 & 3.67 \\
\bottomrule
\end{tabular}
}
\end{table}

\subsection{RQ4: Live execution exposes the enforcement boundary}
\label{sec:live-results}

The live arm asks whether evidence selection remains useful once checks and
writes can interleave. It supports an enforcement boundary rather than a
second demonstration of portfolio risk superiority: the guarded portfolio
completes more tasks than the guarded independent vote, but has slightly
higher observed risk, and the full atomic guard performs best.

Table~\ref{tab:live} reports the preregistered local arm
(Section~\ref{sec:live-method}): \ltEpisodes{} episodes, eight policies, every
effect executed against SQLite, \ltExecutions{} recorded runs.  Ordinary
commits are unsafe on \ltUnsafeNoVerifier\% of episodes.  Cross-model votes on
the shared cache and with an independent read reach \ltUnsafeVoteSame\% and
\ltUnsafeVoteIndep\%; the paired difference (L1, $\ltPOneDiff$
[$\ltPOneLo$, $\ltPOneHi$]) is not significant after Holm correction
($p=\ltPOneP$) over twelve contract clusters, because Phi-4-mini rejects
almost every migration proposal regardless of source, so the shared-cache
vote blocks most migrations and completes only 7\% of them safely.  Under the
common-stale condition alone the shared-cache vote commits four unsafe effects
and the independent read none.

\begin{table}[t]
\centering\small
\caption{Live tool arm: unsafe execution per episode (cluster-bootstrap CI
over 12 contracts; a zero count shows its 95\% upper bound), safe task
completion, and deferral.}
\label{tab:live}
\footnotesize\setlength{\tabcolsep}{2pt}
\begin{tabular}{lrrr}
\toprule
Policy & Unsafe \% [95\% CI] & Safe done \% & Defer \% \\
\midrule
Plain commit & 20.8 [18.5, 23.1] & 79.2 & 0.0 \\
Vote, shared cache & 9.3 [5.1, 13.9] & 48.6 & 0.0 \\
Vote, indep.\ read & 5.6 [3.2, 8.3] & 40.7 & 0.0 \\
Partial guard, preflight & 16.2 [13.4, 19.0] & 80.6 & 0.0 \\
Partial guard, atomic & 13.0 [9.7, 16.2] & 81.0 & 0.0 \\
Indep.\ vote + guard & 2.8 [0.9, 4.6] & 41.2 & 0.0 \\
Portfolio + guard & 4.2 [1.9, 6.5] & 68.5 & 16.7 \\
Full atomic guard & 0.0 [$\le$1.4] & 81.0 & 0.0 \\
\bottomrule
\end{tabular}

\end{table}

The after-check race (C3, 36 episodes) separates verification
from enforcement.  The writer process committed before the action in all 36
episodes of every policy (transaction order recorded in the database log).
Every verifier-only policy then commits the same twelve unsafe effects, since
all evidence was read before the change.  The partial guard as a separate
preflight commits 16 unsafe effects; the same clauses re-checked inside the
write transaction commit 9 (L2, $\ltPTwoDiff$ [$\ltPTwoLo$, $\ltPTwoHi$],
Holm $p=\ltPTwoP$; not significant at 0.05 after correction), and the remaining nine are exactly the revoked approvals
and extra preconditions the partial guard does not express.  The
full-predicate atomic guard commits no unsafe effect in \ltEpisodes{}
episodes (cluster-adjusted 95\% upper bound \ltZeroUpperGuardFullAtomic\%)
and completes \ltSafeGuardFullAtomic\% of tasks safely, tied with the
atomic partial guard for the highest safe completion.

Without the barrier, a random-timing subset uses writer delays of
0, 5, 20, or 50\,ms across 20 episodes per policy (60 runs).
Each tested policy encounters four after-check races and three unsafe effects;
in fourteen episodes the mutation occurs after the action, and in two SQLite
refuses the writer's mutation. Thus the forced-race result describes an
exposed window, not its frequency in an uncontrolled workload. The injector
initially crashed on a refused duplicate-column mutation; it was repaired to
record that outcome, and the stage resumed without dropping rows.

The frozen 5\% portfolio, applied zero-shot through the metadata mapping,
routes fresh contexts to a same-source check followed by the atomic guard,
stale contexts with a healthy $B$ to an independent read, and degraded $B$ to
deferral.  Its unsafe rate is \ltUnsafePortfolio\% (all nine cases in the race
condition) with \ltSafePortfolio\% safe completion and \ltDeferPortfolio\%
deferral, against \ltUnsafeVoteIndepGuardPartialAtomic\% and
\ltSafeVoteIndepGuardPartialAtomic\% for the independent vote with the same
guard (L3 unsafe difference $\ltPThreeDiff$ [$\ltPThreeLo$, $\ltPThreeHi$],
Holm $p=\ltPThreeP$; safe completion $+0.27$ [$+0.13$, $+0.43$]).  The
portfolio buys availability rather than lower risk here and is dominated by
the full guard, the expected boundary when the whole predicate is checkable
in the database.

Lost responses (C4) are handled by the request identifier: idempotent commits
replay the stored response for all 96 retried requests with no duplicate
effect.  Without idempotency, retried snapshot promotions duplicate audit rows
and pointer switches (four unsafe plain commits), whereas duplicate columns
and active attempts are refused by SQLite as tool errors: retry counts do not
equal duplicate effects, and database constraints act as an uncovered guard
for two families.  Under degraded evidence (C5) the portfolio defers every
episode and the independent vote blocks.

Measured gate latency with live verifier calls and one resident model
(layer~B, 72 episodes per policy) shows the cost of model diversity on a
16\,GB machine: tool-only guards decide in milliseconds; the shared-cache
vote, which swaps between two checkpoints, has median \ltLatMedVoteSame\,s
(p95 \ltLatPnfVoteSame\,s), the independent-read vote \ltLatMedVoteIndep\,s
(\ltLatPnfVoteIndep\,s), and the portfolio \ltLatMedPortfolio\,s
(\ltLatPnfPortfolio\,s), since its common path calls one verifier and the
guard; excluding Ollama load time the medians are \ltLatMedNoLoadVoteSame\,s,
\ltLatMedNoLoadVoteIndep\,s, and \ltLatMedNoLoadPortfolio\,s.  Every live
decision matched its cached-verdict counterpart (360/360). A full-workflow
subset runs a live actor and live gate on twelve common-stale episodes for
each of two policies (24 runs). Median end-to-end latency is 19.72\,s
(p95 20.32\,s) for the independent vote and 15.86\,s (p95 16.50\,s) for the
portfolio. Actor medians are 11.48\,s and 11.58\,s, respectively, including
checkpoint loading; these workflow measurements are separate from gate-only
latencies above.

\section{Engineering Implications and Validity}
\label{sec:implications}

\textbf{Use full atomic enforcement when it is available.} If the target
system can evaluate the entire safety predicate within the write transaction,
the live results favor that mechanism: additional model checks offer no
observed safety or completion advantage. Portfolio selection addresses the
remaining case, where guards are partial or unavailable and evidence sources
differ in freshness, competence, and cost. This is an applicability condition
for the controller, not a reason to replace database enforcement.

\textbf{Expose failure lineage to evidence selection.} A
commit gate cannot value redundancy from a source label alone: platforms
should propagate an evidence identifier, upstream lineage, observation time,
schema completeness, and health with every tool result, and checks that share
a lineage should count as one failure domain even across models.  An
independent read, an exact guard, and a person differ in cost, availability,
expressiveness, and common modes; making these explicit permits auditable
routing and makes deferral a normal software outcome.

\textbf{Calibrate to the deployment population.}  The observed 1.9\% test
rate supports the nominal 5\% setting only in this evaluation population; the
approximate calibration is not a formal guarantee, and the fault-family
holdouts show why a target label must not be carried across an unseen failure
mode.  Operational use needs recurring calibration data for each new tool,
lineage, and incident family.

\textbf{Bind checks to execution.}  A live integration should bind a decision
to the exact proposal, resource, evidence version, and policy version, and
invalidate it when any of them changes. The live arm shows why: a guard that reads
its conditions in one transaction and applies the effect in another lets a
concurrent writer slip between them, and no verification before the action
closes that gap; only the check inside the write transaction does, and only
for the clauses it covers.  Request identifiers with idempotent replay turn
lost responses into replayed answers rather than duplicate effects.  These
properties were shown on a loopback service with one writer process, not on
distributed services; lineage discovery and protection against forged
provenance remain outside the implementation.

\textbf{Construct validity.}  Tasks, lag probabilities, telemetry noise, unit
costs, and the live contracts are controlled design choices, so absolute risk
and cost values do not transfer; the supported claim is the ordering of
mechanisms under the stated fault model.  Human deferral is modeled as safe at
fixed cost.

\textbf{Internal and statistical validity.}  Hidden-label audits and frozen
counterfactual replay prevent a controller from seeing scoring fields or
receiving easier proposals; replay is valid because the controller is inserted
after proposal generation, and the replan experiment is reported separately.
The verifier prompt changed once after a training-only pilot.  The template
(or contract) is the sampling unit; only the preregistered comparisons are
multiplicity-corrected, and eight calibration templates or twelve live
contracts cannot support fine risk targets or precise differences.

\textbf{External validity.}  The actor and verifier checkpoints are quantized
4--10B models run on one laptop.  The actors almost never abstain, so the study
primarily evaluates the gate rather than actor calibration.  The external
numeric benchmark fails to reproduce the main effect, which limits the claim
to domains where verifiers can interpret the evidence.  The live arm uses a local service, twelve contracts, and one writer; larger
models, distributed services, correlated human decisions, adversarial faults,
and longitudinal source drift remain open evaluations.

\textbf{Reproducibility.}  The artifact contains prompts, model digests,
generator parameters, salted splits, raw actor/verifier JSONL, calibration
traces, both preregistrations, the live service, writer, and transaction
logs, and scripts that regenerate every reported table from frozen records.
The frozen reproducibility and full-audit archives are publicly available on
Figshare at \url{https://doi.org/10.6084/m9.figshare.33511441.v1}.
Analysis from frozen records requires no model access; the full-audit archive
additionally includes per-episode SQLite databases.

\section{Conclusion}

Reliable agent commit gates require both evidence selection and enforcement.
In the controlled benchmark, source diversity reduces false approval more
than model diversity, and an observable portfolio improves the risk--cost
trade-off over fixed checks. Those benefits are conditional: unseen faults
break calibration transfer, weak verifiers fail to exploit independent
evidence, and better evidence alone does not reliably induce actor correction.

The live study locates a separate limit. Evidence acquired before a concurrent
change cannot enforce the state at execution. Atomic guards protect the
clauses they express, and a full guard is the preferred mechanism when the
entire predicate is checkable. With partial guards, the portfolio improves
completion relative to a guarded independent vote but does not establish
lower risk. Together, these findings support an explicit deployment contract:
record evidence lineage, enforce checkable clauses at commit time, and
calibrate or defer decisions that depend on incomplete evidence. Broader
validation remains necessary for distributed tools and long-horizon agents.

\bibliographystyle{IEEEtran}
\FloatBarrier
\bibliography{refs_se}

\end{document}